\PassOptionsToPackage{numbers,sort&compress}{natbib}
\documentclass[sigconf,nonacm]{acmart}

\usepackage[utf8]{inputenc}
\usepackage[T1]{fontenc}
\usepackage{microtype}
\usepackage{booktabs}
\usepackage{tabularx}
\usepackage{array}
\usepackage{amsmath,amssymb}
\usepackage{graphicx}
\usepackage{xurl}
\usepackage{xcolor}

\setcopyright{none}
\hypersetup{
  hidelinks,
  pdftitle={A Verifier Can Leak the Answer: Diagnosability Before Optimization in Closed-Loop Agent Debugging},
  pdfauthor={Peiying Zhu; Sidi Chang}
}

\providecommand{\tightlist}{\setlength{\itemsep}{0pt}\setlength{\parskip}{0pt}}
\begin{document}
\acmshorttitle{A Verifier Can Leak the Answer}
\acmshortauthors{Zhu \& Chang}
\pagestyle{acmstyle}
\thispagestyle{firstpage}

\twocolumn[
\begin{@twocolumnfalse}
\begin{center}
{\LARGE\bfseries A Verifier Can Leak the Answer: Diagnosability Before Optimization in Closed-Loop Agent Debugging\par}
\vspace{1.0em}

\begin{tabular}{@{}c@{\hspace{2em}}c@{}}
{\large Peiying Zhu\textsuperscript{*}} & {\large Sidi Chang\textsuperscript{*\,$\dagger$}} \tabularnewline
{\small peiying@blossomai.co} & {\small schang@blossomai.co} \tabularnewline
{\small Blossom AI} & {\small Blossom AI Labs} \tabularnewline
{\small San Francisco, CA, USA} & {\small Tokyo, Japan}
\end{tabular}
\vspace{1.2em}
\end{center}

\noindent\parbox{\textwidth}{
\small
\noindent\textbf{Abstract}\\[3pt]
Agent developers increasingly compare prompts, tools, policies, and diagnosis algorithms through simulator-grounded verifiers. A verifier can nevertheless make a solver comparison vacuous: if its probes or predicates encode the target identity, an exact optimizer may appear effective without resolving any genuine ambiguity. We report such a failure in an aggregate-trace debugger for a closed-loop decision agent. Exact minimum hitting set (MHS) and a propagation-aware greedy method returned identical supports in 12/12 development cases and the same planted-fault recovery in 9/12. A subsequent audit found that exact-anchor predicates produced the planted pair in 9/9 cases. After removing those anchors, overall planted-pair recovery was 8/9; hard-probe singleton pairs nevertheless matched the planted pair in 9/9, and no case retained a nonempty residual conflict family after propagation (0/9). The optimizer was correct, but the verifier had already disclosed the answer. We replace solver-first evaluation with a support-gated verification contract. A clean reference map must first show repeated component exposure; a matched reference/current gate must then establish comparable runtime evidence; only afterward may an independently calibrated signal rule return a detection. In a preregistered heldout comprising 1,440 cases and 21,600 partition rows, 55/72 regime-component units passed the reference gate, 54/55 passed the runtime gate, and stable false admission was 0/20 represented components with a one-sided exact 95\% upper bound of 0.1391. Within admitted units, affected clean traffic predicted detection better than nominal fault-cell fraction. The main lesson is structural: verify evidence eligibility and non-revelation before optimizing the component selector. Otherwise a stronger solver can merely certify a stronger verifier artifact.

}

\vspace{1.5em}
\end{@twocolumnfalse}
]

\renewcommand{\thefootnote}{\fnsymbol{footnote}}
\footnotetext[1]{Both authors contributed equally to this research.}
\footnotetext[2]{Corresponding author.}
\renewcommand{\thefootnote}{\arabic{footnote}}

\section{Introduction}\label{introduction}

Interactive-agent benchmarks have moved evaluation from static answer matching toward execution in software, web, database, and tool environments \cite{manualRef01,manualRef02,manualRef03,manualRef04}. This is progress: an agent should be judged by what happens in its environment, not only by what it says. It also creates a second-order problem. The environment, probe design, state abstraction, and verifier jointly determine what evidence reaches the score. When any of these mechanisms is wrong, an optimizer can improve the reported result without improving the underlying agent or diagnosis.

This paper studies that problem in a deliberately controlled setting. Our system is a closed-loop decision agent, not a language model. It selects actions under changing demand and inventory conditions. We observe only aggregate traces and try to localize a planted fault to one or more policy components. The diagnostic pipeline compiles aggregate discrepancies into conflicts and then chooses a low-cost component set that intersects every conflict. Exact MHS is a natural downstream optimizer \cite{manualRef05,manualRef07}. The original development claim was that exact MHS should improve fault localization over a weighted greedy alternative.

That claim did not survive an audit. The two solvers agreed because the compiled conflict family was already solved. An exact-component predicate and a one-component hard probe interacted to create planted-component singletons. Once singleton propagation ran, neither exact nor greedy selection had a substantive choice. This is a verifier failure, not a mathematical failure of MHS.

The failure suggests a more general principle for agent development:

\begin{quote}
A verifier must establish that its evidence is eligible and non-revealing before its score can support an optimizer comparison.
\end{quote}

``Eligible'' means that the deployed policy actually visited the component often enough for aggregate traces to carry information, and that reference and current observations are comparable. ``Non-revealing'' means that the evaluation intervention does not encode the hidden answer through its scope or mechanics. These conditions are distinct. More traffic cannot repair answer leakage, and a leakage-free predicate cannot create evidence for a component the agent never exercised.

We make four contributions.

\begin{enumerate}
\def\labelenumi{\arabic{enumi}.}
\tightlist
\item
  We present a concrete verifier-leakage audit. Exact MHS and greedy selection agree not because the optimization problem is easy in a meaningful sense, but because probes and predicates collapse the conflict family into forced singletons.
\item
  We define a two-stage diagnosability gate for aggregate-trace verification: a clean reference-map gate followed by a matched runtime two-stream gate. Unsupported units abstain instead of receiving a diagnostic score.
\item
  We independently calibrate stable false admission at the physical-component level, separate from coverage and detection power.
\item
  We execute a preregistered heldout and show that the gated verifier is operational: it admits 55/72 units at reference, rejects one additional represented component at runtime, and produces no stable false admissions among 20 represented components. Within admitted cases, affected clean traffic is a better power coordinate than structural fault size.
\end{enumerate}

The heldout does not compare exact MHS with greedy selection and does not establish localization accuracy. That omission is intentional. Once the development audit showed that the evidence construction did not pose a genuine residual optimization problem, a new solver contest would have repeated the same mistake. The confirmatory experiment instead tests whether a verifier can say when its own evidence is usable.

\section{From environment-grounded evaluation to verifier-grounded claims}\label{from-environment-grounded-evaluation-to-verifier-grounded-claims}

AgentBench evaluates language-model agents across several interactive environments \cite{manualRef01}. WebArena emphasizes reproducible, functional web environments \cite{manualRef02}. SWE-bench uses repository tests to judge code-changing agents \cite{manualRef03}, while \(\tau\)-bench compares final database states and repeated-trial reliability in tool-agent-user interactions \cite{manualRef04}. These benchmarks illustrate why environment grounding matters: completion text is often a poor proxy for whether an action sequence achieved the requested state.

Yet environment grounding does not automatically validate a verifier. A unit test may omit a requirement; a simulator may expose privileged state; a judge may reward a shortcut; a probe may alter the object it purports to measure. The relevant question is not only whether the verifier is deterministic or reproducible, but whether its operationalization supports the intended claim.

Model-based diagnosis (MBD) provides a useful decomposition. Conflicts are first derived from a system model and observations; candidate diagnoses are then computed as hitting sets of those conflicts \cite{manualRef05}. Modern exact search can improve memory and runtime without changing this conditional structure \cite{manualRef07}. Classical GDE also separated diagnosis from sequential measurement selection, using expected entropy to choose measurements that discriminate among candidates \cite{manualRef06}. Diagnosability research similarly asks whether observations can distinguish faults before celebrating a diagnoser \cite{manualRef08}.

Our setting differs from passive GDE measurements in one crucial way: the audited hard probe is a counterfactual intervention on a single component. That intervention can be useful, but it also creates a leakage channel when paired with a predicate scoped to the same component. The agent-verification analogue is a benchmark tool that writes the hidden label into an observable field and then rewards the agent for reading it.

Abstention supplies the complementary idea. Selective prediction treats refusal as a valid output when confidence or coverage is inadequate \cite{manualRef09}. Risk-controlling calibration can attach finite-sample guarantees to a chosen loss \cite{manualRef10}. We apply the same discipline upstream: the verifier abstains before diagnosis when the policy did not produce enough matched evidence. This is not a confidence threshold on a completed prediction; it is an eligibility test on the evidence from which a prediction would be constructed. Dataset-shift tests and conformal change detectors provide a related lifecycle pattern: detected reference mismatch should trigger review or retraining rather than silently extending the original claim \cite{manualRef11,manualRef12}.

\section{Closed-loop debugging setting}\label{closed-loop-debugging-setting}

\subsection{Agent and observable components}\label{agent-and-observable-components}

The agent is a fitted closed-loop target policy operating in a simulator. Its 24 disjoint components are indexed by time quarter, inventory half, and market third. Three separately fitted demand regimes, \(\lambda_0\in\{5,7,9\}\), produce 72 regime-component units. The diagnostic observes typed aggregate traces rather than the planted component identity.

For each evaluation partition, the signal calculation uses a regional distribution distance and a mean reference-current action gap. A unit is assessed over 15 independent partitions. The policy is refitted per regime for the formal study, isolating evidence behavior from cross-regime policy mismatch.

\subsection{Fault interventions and exposure}\label{fault-interventions-and-exposure}

A fault shifts a deterministic subset of direction-changeable target-field cells by one action bucket. Every selected subset is tested in outward and inward directions. Two mask families are frozen:

\begin{itemize}
\tightlist
\item
  uniform, a seeded random ordering of changeable cells; and
\item
  flow-weighted, a seeded weighted-without-replacement ordering favoring frequently visited clean cells.
\end{itemize}

Nominal cell fraction is a structural description of the intervention. It need not describe what the agent experienced. Let \(o_c\) be the clean visit count for component cell \(c\) and \(S_d\) the cells selected in direction \(d\). Directional affected traffic is

\[
\tau_d=\frac{\sum_{c\in S_d}o_c}{\sum_{c\in \text{component}}o_c},
\]

and the primary bidirectional exposure is \(\tau=(\tau_{out}+\tau_{in})/2\). If clean occupancy is zero, \(\tau=0\).

\subsection{Conflict compilation and downstream optimization}\label{conflict-compilation-and-downstream-optimization}

The development diagnostic converts aggregate evidence and scoped predicates into conflict sets. Given component costs, exact MHS returns a minimum-cost component set intersecting every conflict. Weighted greedy selection repeatedly chooses a cost-effective component and performs the same singleton propagation.

The exact guarantee is real but conditional: it optimizes the compiled conflict family. It does not establish that the conflict family preserves physical ambiguity, that a planted component is recoverable, or that exact search improves over greedy selection on the instances actually produced by the verifier.

\section{Failure audit: the verifier had already solved the task}\label{failure-audit-the-verifier-had-already-solved-the-task}

Table 1 records the development evidence that forced us to withdraw the optimizer-advantage claim. These are audited development results, not formal heldout endpoints.

The original instances were not devoid of combinatorial structure: non-singleton conflicts comprised 55.5\% of the conflict collection, and every one of the 12 scenarios contained overlapping conflicts. Yet exact MHS and weighted greedy still returned identical supports in all 12 scenarios.

\begin{table*}[t]
\caption{Development audit of the original MHS claim.}
\centering
\small
\begin{tabularx}{\textwidth}{@{}>{\raggedright\arraybackslash}p{0.34\textwidth}c>{\raggedright\arraybackslash}X@{}}
\toprule
Audit & Finding & Consequence \\
\midrule

Exact MHS vs weighted-greedy support equality & 12/12 & no observed decision advantage \\
Exact planted recovery, exact MHS & 9/12 & good recovery, not superiority \\
Exact planted recovery, weighted greedy & 9/12 & matched exact MHS \\
Exact-anchor singleton pairs match planted pair & 9/9 & anchor layer reveals the target \\
Recovery after removing exact anchors & 8/9 & deleting one shortcut is insufficient \\
Hard-probe singleton pairs match planted pair & 9/9 & probe-scope interaction still reveals target \\
Nonempty residual conflicts after propagation & 0/9 & no residual optimizer contest remains \\
\bottomrule
\end{tabularx}
\end{table*}

The mechanism is simple. The predicate asks about exactly one named component. The hard probe intervenes on exactly that component. The local response is converted into a singleton conflict naming the planted component. Removing an explicit anchor rule does not remove this second channel. After propagation, both solvers inherit the same forced choice.

This diagnosis changes what ``verification'' means. A conventional audit might check that the MHS implementation returns a minimum-cost hitting set and that the greedy baseline is implemented correctly. Both checks can pass while the scientific comparison remains empty. A stronger audit must ask:

\begin{enumerate}
\def\labelenumi{\arabic{enumi}.}
\tightlist
\item
  Did the verifier\textquotesingle s upstream transformations encode hidden target identity?
\item
  After deterministic propagation, does a nontrivial residual decision remain?
\item
  Is the agent-generated trace supported for the component under test?
\item
  Are the reference and current traces comparable on the same partitions?
\end{enumerate}

Only the first two questions concern non-revelation; the last two concern evidence eligibility. We found that both classes of checks are needed.

\section{A support-gated verification contract}\label{a-support-gated-verification-contract}

\subsection{Reference-map gate}\label{reference-map-gate}

A component has support in a clean partition when its existing support count is at least 12. A regime-component unit enters the reference map when at least 14 of 15 clean partitions have support. All 72 units are screened; no component is excluded by identity or by its development behavior.

This gate asks whether the closed-loop agent ever generated enough clean evidence about a component. It does not use planted fault identity and does not infer a diagnosis. A failed unit returns \texttt{REFERENCE\_ABSTAIN}.

\subsection{Runtime two-stream gate}\label{runtime-two-stream-gate}

At runtime, a partition is usable only when both its reference stream and its current stream have support of at least 12 in that same partition. Admission requires at least 14 jointly supported partitions out of 15. Bidirectional observations require both directions to pass. A failed unit returns \texttt{RUNTIME\_ABSTAIN}.

The intersection is essential. Separate totals can both look adequate while referring to different partitions. Matched support ensures that the aggregate difference is grounded in comparable evidence.

\subsection{Stable signal and independent null calibration}\label{stable-signal-and-independent-null-calibration}

For each matched partition, signal strength is \(z=\max(\mathrm{region\_d1}/0.20, |\mathrm{mean\_action\_gap}|/0.35)\). The cutoff \(z\geq1.50\) was selected before formal execution from the fixed grid \(\{1.00,1.25,1.50,1.75,2.00\}\) and was never retuned on the formal null. A direction is stable when at least 14 of 15 matched partitions signal, and a unit-level detection requires both directions to be stable.

Stable false admission is calibrated on a separate clean-null schedule with disjoint seeds. The primary sampling unit is the distinct physical component, not the regime-component row. The frozen safety statement is intentionally modest: the one-sided exact 95\% upper confidence bound on component-level stable false admission must be at most 0.20. It is not a 5\% false-alarm guarantee.

\subsection{Result state and claim separation}\label{result-state-and-claim-separation}

The verifier state is

\[
\begin{gathered}
\texttt{REFERENCE\_ABSTAIN} \rightarrow \texttt{RUNTIME\_ABSTAIN} \\
\rightarrow \texttt{SIGNAL\_ELIGIBLE} \\
\rightarrow \{\texttt{DETECTED},\texttt{NOT\_DETECTED}\}.
\end{gathered}
\]

These are conditional states, not a requirement that every unit reach a score. We also separate three decision layers:

\begin{itemize}
\tightlist
\item
  \textbf{execution:} whether the frozen protocol and independent validation were executed;
\item
  \textbf{operational safety:} whether the false-admission bound met its frozen criterion; and
\item
  \textbf{structural hypotheses:} whether traffic beats cell fraction, traffic is practically sufficient across the two frozen mask families, and aggregate power is nondecreasing.
\end{itemize}

There is no combined scientific PASS/FAIL label. This prevents a development team from treating one favorable endpoint as permission to ignore an unfavorable one---or from changing the verifier because an unrelated endpoint missed its threshold.

\section{Preregistered heldout}\label{preregistered-heldout}

The formal schedule contains 1,440 cases, 21,600 partition rows, and 3,456,000 simulator episodes. An episode is a Monte Carlo trajectory, not an independent statistical unit. Coverage begins with 72 regime-component units; safety and bootstrap inference use 20 represented physical-component clusters; and the structural fit contains 540 unit-arm rows nested within those clusters. Development, pilot, and formal seeds are disjoint; seeds 204800--204999 remain unused. The run locks the protocol, code, configuration, environment, schema, and validator to a git commit. Reference-gate failures remain scheduled so that schedule validation is independent of observed eligibility.

The formal analysis has three families of endpoints.

\begin{enumerate}
\def\labelenumi{\arabic{enumi}.}
\tightlist
\item
  \textbf{Coverage and safety:} reference admission, runtime rejection, and component-level stable false admission.
\item
  \textbf{Traffic-indexed power:} five traffic targets in each mask family, with conditional and operational detection counts.
\item
  \textbf{Structural tests:} clustered comparisons of traffic versus cell fraction and of a traffic-only model versus a family-plus-interaction model.
\end{enumerate}

The confirmatory experiment does not compare MHS and greedy selection. It tests the replacement verifier\textquotesingle s admission, calibration, and signal behavior.

\section{Results}\label{results}

Independent validation passed all 56 formal checks. The protocol was therefore labeled \texttt{EXECUTED}; the scientific endpoints retained separate labels.

\subsection{The verifier refuses unsupported units}\label{the-verifier-refuses-unsupported-units}

The clean reference gate admitted 55 of 72 regime-component units (76.4\%), representing 20 of 24 physical components. Among represented components, reference-map invalidation was 0/20. The runtime two-stream gate rejected one of those 20 components; at the regime-component level, 54 of 55 reference-admitted units remained runtime-admitted.

These abstentions are not missing data to be repaired by imputation. They describe where the deployed policy failed to generate the evidence needed for aggregate diagnosis.

\subsection{Stable false admission met the frozen safety statement}\label{stable-false-admission-met-the-frozen-safety-statement}

The independent clean-null arm produced 0 stable false admissions among 20 represented components. The one-sided exact 95\% upper bound was 0.1391, below the frozen 0.20 threshold. The operational label was therefore \texttt{SAFETY\ CONFIRMED}. At the regime-component level, 0/54 admitted units were false stable, with a descriptive upper bound of 0.0540.

The zero count does not establish that the true rate is zero. The upper bound is the claim.

\subsection{Agent traffic, not edit size, governs verifiability}\label{agent-traffic-not-edit-size-governs-verifiability}

At the five traffic targets, bidirectionally stable detection counts were nondecreasing in both mask families:

\begin{itemize}
\tightlist
\item
  uniform: 0, 17, 37, 52, 54;
\item
  flow-weighted: 0, 13, 35, 49, 54.
\end{itemize}

A clustered logistic comparison showed that replacing affected clean traffic with nominal selected-cell fraction increased negative log likelihood by 0.1264 nats per unit-arm row; the two-sided 95\% component-cluster interval was {[}0.0593, 0.1918{]}. Traffic was therefore the better primary power coordinate within admitted formal units. Adding mask family and traffic-by-family interaction improved log loss by only 0.0015 nats per unit-arm row, with a one-sided 95\% upper bound of 0.0066, meeting the frozen 0.01-nat practical-sufficiency rule within the two frozen families. However, the secondary descriptive traffic-plus-family coefficient was -0.3618 with interval {[}-0.8617, 0.0367{]}, corresponding to an odds-ratio interval {[}0.42, 1.04{]}. Because that interval extends below the frozen practical-equivalence range {[}0.5, 2.0{]}, it does not establish coefficient-scale equivalence; the \texttt{PASS} applies only to the log-loss sufficiency endpoint.

This result matters for verification design. A counterfactual edit can touch many cells the agent never visits and remain nearly invisible, or touch few high-traffic cells and become easy to detect. Structural intervention size is not evidence exposure.

\begin{table*}[t]
\caption{Compact formal verifier outcome.}
\centering
\footnotesize
\setlength{\tabcolsep}{3pt}
\renewcommand{\arraystretch}{1.05}
\begin{tabularx}{\linewidth}{@{}>{\raggedright\arraybackslash}p{0.10\linewidth}>{\raggedright\arraybackslash}p{0.24\linewidth}>{\raggedright\arraybackslash}p{0.27\linewidth}>{\raggedright\arraybackslash}X@{}}
\toprule
Layer & Endpoint & Formal result & Interpretation \\
\midrule
execution & protocol, hashes, schedule, validator & 56/56 checks; 1,440 cases & \texttt{EXECUTED} \\
evidence & reference admission & 55/72 units; 20/24 components & unsupported units abstain \\
evidence & runtime admission & 54/55 reference-admitted units & one additional runtime abstention \\
safety & component stable false admission & 0/20; upper 0.1391 & \texttt{SAFETY\ CONFIRMED} under 0.20 rule \\
structure & traffic versus cell fraction & 0.1264 {[}0.0593, 0.1918{]} nats/row & traffic is better within admitted units \\
structure & aggregate traffic-target power & monotone counts in both families & structural hypothesis \texttt{PASS} \\
\bottomrule
\end{tabularx}
\end{table*}

\section{Implications for reliable agent development}\label{implications-for-reliable-agent-development}

\subsection{Verify the compiler, not only the scorer}\label{verify-the-compiler-not-only-the-scorer}

Agent evaluation pipelines often include hidden compilation steps: converting trajectories to events, events to predicates, predicates to rewards, or tool states to success labels. A final scorer can be formally correct while these upstream transformations leak the target or erase ambiguity. Red-teaming should therefore include target-swapping, scope-mismatch, and residual-ambiguity tests.

A practical invariant is: after removing forced facts and deterministic propagation, the verifier should report whether a nontrivial decision remains. If the residual candidate family is empty or singleton in every case, a solver benchmark cannot support claims about optimizer quality.

\subsection{Separate intervention from measurement}\label{separate-intervention-from-measurement}

Passive observation, active information gathering, and counterfactual intervention are not interchangeable. Classical measurement selection asks which observation would reduce uncertainty \cite{manualRef06}. An intervention can change the system and may encode its target in the response. Agent benchmarks that expose database state, execute privileged tools, or inject faults should document which of these roles each probe plays.

\subsection{Treat abstention as a verifier output}\label{treat-abstention-as-a-verifier-output}

Environment-grounded benchmarks frequently force every task into success or failure. In closed-loop settings, the agent may never enter the state region needed to test a particular requirement. Reporting failure in that case confounds lack of capability with lack of evidence; reporting success is worse. A support gate makes the denominator visible and turns ``not testable here'' into a first-class outcome.

\subsection{Keep claims modular}\label{keep-claims-modular}

Execution integrity, low false admission, adequate coverage, monotone power, and solver superiority are different claims. Collapsing them into one benchmark score creates opportunities for selective interpretation. A claim ledger makes regressions local: an agent update may preserve verifier safety while reducing coverage, or improve traffic exposure while leaving diagnosis quality unchanged.

\paragraph{Reproduction package.} The reproducibility artifact is available at \url{https://anonymous.4open.science/r/artifact-9f37d2/verify_agents_2026/README.md}. It includes the frozen protocol and code, simulator source, all saved reference/mask/case/partition rows, and an anonymization map. From its unpacked root, \texttt{python3 verify.py} checks the content manifest and independently reconstructs all 56 validation checks, exact bounds, model fits, bootstrap draws, and decision labels without generating trajectories.

\section{Limitations}\label{limitations}

This is one simulator, one closed-loop policy family, three demand regimes, and 24 disjoint components. The agent is not an LLM, and we do not claim that the exact support thresholds transfer to language-agent benchmarks. The contribution is a verification structure and a falsified optimizer claim, not a universal threshold.

The formal heldout confirms admission, false-admission calibration, and traffic-indexed signal behavior. It does not confirm planted-fault localization and does not evaluate genuinely ambiguous residual conflict families. Exact MHS remains a valid risk-cost optimizer when such families exist; this study shows only that the audited verifier failed to produce them.

The family-sufficiency bound is based on approximately 20 represented component clusters. A percentile cluster bootstrap at that cluster count may undercover in a direction favorable to the result. The 0.0066 upper bound has only a 0.0034-nat margin below its frozen threshold.

The refresh endpoint is a state-machine smoke test. It reuses the same frozen reference buffer and therefore cannot show that refresh repairs a stale map after real drift.

\section{Conclusion}\label{conclusion}

The most dangerous verifier is not necessarily noisy. It can be deterministic, reproducible, and mathematically correct while making the scientific comparison vacuous. In our closed-loop debugger, exact-scope predicates and one-component interventions produced forced planted-component singletons, so exact MHS never faced the ambiguity it was supposed to resolve.

We replaced that solver-first claim with a support-gated contract that verifies whether the agent generated comparable evidence before returning a signal. The formal heldout shows that this contract can operate with explicit abstention, independently calibrated false admission, and traffic-indexed power. The next optimizer comparison should begin only after a redesigned verifier demonstrates nonempty residual ambiguity. Reliable agent development requires verifying the evidence generator before trusting the algorithm that optimizes its output.

\bibliographystyle{plain}
\bibliography{references}

\end{document}